\documentclass[prb,twocolumn]{revtex4-1} 

\usepackage[breaklinks,colorlinks,citecolor=blue]{hyperref}

\usepackage{graphicx}

\usepackage[utf8]{inputenc}

\usepackage{amsfonts}
\usepackage{amssymb}

\newcommand{\be}{\begin{equation}}
\newcommand{\ee}{\end{equation}}
\newcommand{\bea}{\begin{eqnarray}}
\newcommand{\eea}{\end{eqnarray}}
\newcommand{\Dd}{\mathrm{d}}

\usepackage{etoolbox}
\usepackage[dvipsnames]{xcolor}

\begin{document}

\title{Detecting gravitational waves with light}

\author{Markus P\"ossel}
\email{poessel@hda-hd.de} 
\affiliation{Haus der Astronomie and Max Planck Institute for Astronomy, K\"onigstuhl 17, 69124 Heidelberg, Germany}


\date{\today}

\begin{abstract}
The strong evidence for low-frequency gravitational waves from pulsar timing arrays (PTAs), published in 2023, has widened the scope for teaching about gravitational wave astronomy. This article provides a simple, unified overview of the detection of gravitational waves using light waves that encompasses the recent PTA detections, the by-now classic interferometric detections using LIGO and similar detectors, and the yet-to-be-accomplished detections using long-arm detectors like the spaceborne LISA. The presentation is at a level accessible for undergraduate students. The influence of gravitational waves on light is derived in a way that makes use only of basic gravitational wave properties and Einstein's equivalence principle.\footnote{The version of record for this article has been published in the {\em American Journal of Physics}, 93, 499--510 (2025), \doi{10.1119/5.0228933}. This present version represents the state of the article after peer review and editorial feedback.}
\end{abstract}

\maketitle 

\section{Introduction}
Results published in the summer of 2023 provide strong evidence for a low-frequency gravitational wave background from measurements using pulsar timing arrays (PTA).\cite{PulsarTiming2023a,PulsarTiming2023b,PulsarTiming2023c,PulsarTiming2023d} In January 2024, the European Space Agency gave the go-ahead for the space-based gravitational wave detector LISA, slated for launch in 2037.\cite{ESA2024} Both developments provide a challenge for teaching about gravitational waves at an undergraduate level: Neither PTA nor LISA can be understood using the so-called short-arm approximation for interferometric detectors that is commonly employed when teaching about gravitational wave detectors.\cite{Spetz1984,Farr2012,Choudhary2018,vanHeijningen2021} 

The aim of this article is to provide teachers with a comprehensive account of gravitational wave detection with electromagnetic radiation, i.e. light, which encompasses both pulsar timing and a setup like that of LISA, but also detectors like LIGO, at the level of introductory physics or astronomy courses. After a review of the basics of gravitational waves in Sec.~\ref{sec:basic-gw}, we deduce their influence on light propagation in Sec.~\ref{sec:GW-light}. We then apply the results to different kinds of detection scenarios, demonstrating that all of them can be understood along the same basic principles: spacecraft transponders in Sec.~\ref{sec:transponders}, pulsar timing in Sec.~\ref{sec:PTA} and interferometric detectors in Sec.~\ref{sec:interferometricDetectors}. 

\section{Gravitational wave basics}
\label{sec:basic-gw}
\begin{figure}[h]
\includegraphics[width=\linewidth]{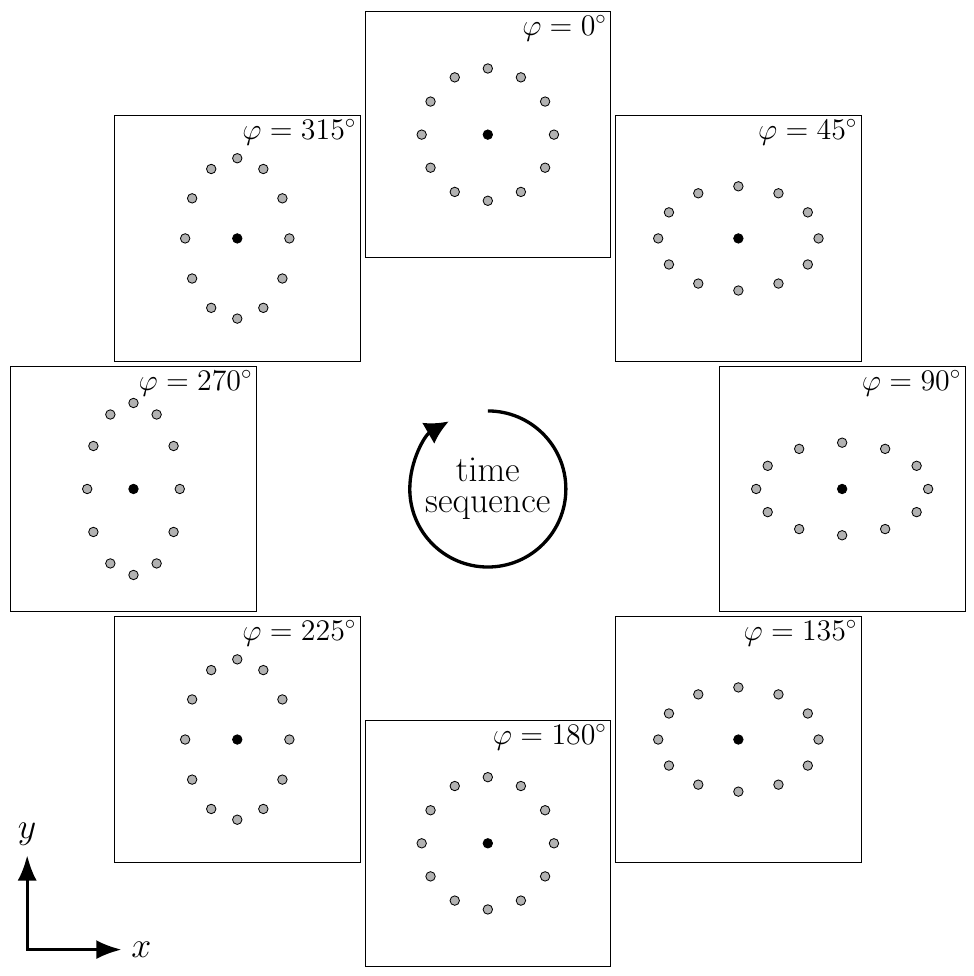}
\caption{
Action of a linearly polarized, purely sinusoidal gravitational wave, propagating orthogonally to the figure plane in the $z$ direction, on a circle of free-floating test particles. Shown are eight snapshots of the time evolution of distances around a central test particle. The time sequence progresses clock-wise. Each snapshot shows the same region of the $x$-$y$ plane
\label{fig:gw-action}
}
\end{figure}

Visualizations of gravitational waves typically feature the basic quadrupole pattern shown in Fig.~\ref{fig:gw-action}, where stretching of distances in one direction on the plane always coincides with shrinking in the orthogonal direction, and vice versa. %
The pattern illustrates the effect of a passing gravitational wave on free-floating test particles, here arranged in a circle with one particle in the center. The wave in question is a {\em plane wave}: We can imagine three-dimensional space as ``sliced up'' into parallel planes, which are orthogonal to the gravitational wave's direction of propagation (here chosen to be the $z$ direction), with the phase of the gravitational wave the same within each slice. Like their electromagnetic counterparts, gravitational waves are transverse: test particle accelerations are orthogonal to the direction of propagation. That the pattern of stretching and shrinking is strictly separated in the $x$ and $y$ directions (instead of changing direction over time) makes this particular example a {\em linearly polarised} wave.

For the following, let us concentrate on one of the parallel planes, say: the plane $z=0$. Characteristically, test-particle distances within that plane change by a (direction-specific) {\em factor}: If two particles are initially separated in a given direction by the distance $L$, a passing gravitational wave will change that distance in proportion to some factor $a(t)$, as $a(t) \cdot L$. All separations between free-floating particles in the same direction will vary by the same factor; an initial distance $2L$ will vary as $a(t)\cdot 2L$, and so forth. For the gravitational wave shown in Fig.~\ref{fig:gw-action}, distances in the $x$ direction are changed by a factor $a_x(t)$; distances in the $y$ direction by another factor $a_y(t)$. As one would expect, changes in length that have both an $x$ and a $y$ component can be calculated using Pythagoras's theorem. The fact that this is a so-called quadrupole pattern can be expressed by $a_x(t)=1/a_y(t)$.

The extreme weakness of gravitational waves reaching the Earth makes it convenient to write  
\be
a_x(t) = 1+\frac12 h(t),
\label{eq:SimpleFormAandH}
\ee
with a dimensionless function $h(t)$, the {\em gravitational wave strain}, which satisfies $|h|\ll 1$. The factor $1/2$ is conventional, chosen for consistency with the usual linearized description of gravitational waves in general relativity. The strain encodes the relative length change 
\be
\frac{\Delta L}{L}=\frac12 h.
\label{eq:StrainDef}
\ee
Since $|h|\ll 1$, we will routinely discard higher-than-linear terms in $h$. Notably, we can write
\be
a_y(t)=\frac{1}{a_x(t)}=\frac{1}{1+\frac12 h(t)}\approx 1-\frac12 h(t).
\label{eq:SimpleFormAandHY}
\ee 
The specific wave in Fig.~\ref{fig:gw-action} is sinusoidal, $h(t) = h_0\sin(\omega t)$.

If we want to describe the action of this particular gravitational wave in another of the parallel planes, at different value of the $z$ coordinate, the generalisation is simple: All we need to do is replace the time variable $t$ by the delayed time $t-z/c$, and consider $a_x(t-z/c)$ instead of $a_x(t)$, and analogously for $a_y$ and for $h$. This construction shows clearly that our wave is propagating at the speed of light $c$ in the positive $z$ direction.

Before we can examine the influence of our gravitational wave on light propagation, we need to give some thought to suitable coordinates. Imagine that our plane is densely filled with freely-floating test particles, all at rest relative to each other prior to the arrival of the gravitational wave. Each particle's world line has an associated proper time: duration as measured by a co-moving ideal clock. Before the gravitational wave makes itself felt, we assume spacetime geometry to be governed by special relativity, and we synchronise the co-moving clocks accordingly. Once the gravitational wave has arrived, we continue to assign to each event $\cal E$ in our plane the time $t$ shown by the co-moving clock of the free-floating particle that is present at $\cal E$. 

Similarly, before the gravitational wave arrives, we assign to each of our family of free-floating particles Euclidean $x,y$ coordinates corresponding to its position, with the central particle in our circle as the spatial origin. We keep those coordinates for each particle fixed (``comoving coordinates'') even during the passage of the gravitational wave, and to each event $\cal E$ in our plane, we assign the coordinates $x,y$ of the unique free-floating particle from our family that is present at $\cal E$. Inter-particle distances calculated via Pythagoras's theorem in these co-moving coordinates only correspond to physical distances in the absence of gravitational waves. In the presence of gravitational waves, we need to multiply $x$ and $y$ coordinate differences with $a_x(t)$ and $a_y(t)$, respectively, to obtain physical distances.

In the general-relativistic formalism for describing small-amplitude gravitational waves,
this coordinate choice, with the time coordinate defined via co-moving clocks and spatial coordinates via distances at a given reference time, is known as the ``TT gauge.'' It is one of several possible gauge choices for describing linearized gravitational waves, and the physical effects on any of the detector configurations presented in the following are of course independent of the chosen gauge. For the purposes of this article, the TT gauge has the considerable advantage that it allows  for a derivation of gravitational-wave effects on light that requires little more than Einstein's equivalence principle: the fact that even within a gravitational field, physics in an infinitesimal spacetime region around an object that is in free fall is governed by the laws of special relativity. This allows for a derivation of the influence of gravitational waves on light that is suitable for students who are not familiar with the basic formalism of general relativity.
 
\section{How gravitational waves influence light}
\label{sec:GW-light}

\subsection{Modelling Light}
\label{sec:DefLight}

Before we examine the influence of gravitational waves on light, let us make explicit what we mean by light in this context. Following standard usage in astrophysics,\cite{Ryden} we use  ``light'' to refer to all varieties of electromagnetic radiation, not just the more specific ``visible light.'' Electromagnetic radiation, in turn, is a quantum phenomenon, reaching our detectors and interacting with optical elements such as mirrors as a stream of photons. For the practicalities of gravitational wave detection, those quantum properties play an important role. In detectors like LIGO, the fact that light is reflected at mirrors not as a smooth and continuous energy flow, but as the stochastic rat-a-tat of photons, is responsible for part of the noise that makes gravitational wave signals hard to detect. This ``quantum noise,'' together with the thermal noise associated with thermal fluctuations, defines the ``noise floor'' that fundamentally limits the sensitivity of a detector design.\cite{LSC2015} The latest generation of ground-based detectors goes so far as to use so-called ``squeezed light,''\cite{Bauchrowitz2013} manipulating the quantum properties of radiation in a way that suppresses the associated noise in a way that cannot be described by classical physics. Such noise estimates, however, and even more so nonclassical light, are beyond the scope of the present article. 

Furthermore, even for the classical electromagnetic field, we do not require the full description in terms of electric and magnetic field vectors. Instead, we model interference effects in a simplified way that is commonly used when teaching about the basics of interference and interferometry:\cite{FeynmanLectures} electromagnetic waves are modelled as scalar waves, usually taken to be sinusoidal, characterised at each location by a value for the displacement and a phase. For such a simplified wave, the displacement can be interpreted as representing the component of the electric field vector in the direction of polarisation for a plane, linearly polarized electromagnetic wave. In that last respect, at least, the model is rather close to reality: Plane waves of this kind are particularly suitable for interferometry, and in detectors like LIGO, considerable technical effort is invested in creating electromagnetic waves with just the right properties: linearly polarized, high-intensity, singular-frequency, fundamental-mode (which roughly translates to: plane-wave-like) laser light.

In situations where our analysis does not require the wave properties of light, we will resort to an even simpler picture: We will model light pulses as point particles travelling at the speed of light, using the basic picture common in special and general relativity where light propagation is described in terms of ``light-like'' world lines. Note that, on the quantum level, there is no fundamental difference between a light pulse (essentially, a bunch of photons) and the travelling maximum of an electromagnetic wave (again, essentially, a bunch of photons). In our model, light pulses and the maxima (or any other fixed-phase points) of a sinusoidal wave travel at the same speed, in special relativity: at the usual speed of light $c$.

\subsection{Equivalence principle and light propagation}
\label{sec:EquivLight}

Consider a light pulse propagating in the $x$ direction in one of our transverse planes. If we were in special relativity (or classical physics), we would automatically assume that a pulse that starts out in the $x$ direction will keep propagating in the $x$ direction. In general relativity, where light gets deflected under the influence of gravity, this statement should not be taken for granted, but in the special situation we have here, symmetries guarantee that our light pulse indeed keeps propagating in the $x$ direction: The gravitational wave's effects are transverse, so we know our pulse will not deviate out of the plane. And by construction, the gravitational wave's effects are symmetric about the $x$ axis, so none of the directions in which the trajectory of our pulse could deviate {\em within} the $x$-$y$-plane is preferred relative to the others. So even while we are not in the gravity-free realm of special relativity, the set-up for our simple gravitational wave ensures that a light pulse that starts out in the $x$ direction will continue to propagate in the $x$ direction (and an analogous statement holds for a light pulse propagating in the $y$ direction).

Wherever the light pulse passes, there will be one of the family of free-floating particles that we introduced in Sec.~\ref{sec:basic-gw}.
At this point, we make use of Einstein's equivalence principle, which encapsulates a realization Einstein had at the very beginning of his path to his theory of general relativity, and that he later called the happiest thought of his life: For an observer in free fall in a gravitational field, the immediate effects of gravity --- such as the pull felt by an observer standing on the Earth's surface --- are absent. Since locally, all objects fall at the same rate, such an observer would see other objects floating alongside themselves. Enclose the observer in a small cabin, and they would not be able to distinguish whether they were in free fall in a gravitational field, or else in deep space, far from all sources of gravity. Einstein generalized this statement to encompass all of physics, and the result is known as Einstein's equivalence principle: for an observer in free fall in a gravitational field, the local physical laws are those of special relativity --- at least in an infinitesimally small neighbourhood of space-time. 

For our space-filling family of free-floating particles, this means we can proceed as follows. We had introduced comoving coordinates $x$ and $y$, defined via physical distances in the absence of any gravitational wave. For a free-floating particle on our $x$ axis, its (permanently assigned) $x$ coordinate is its distance from the origin in the absence of gravitational waves. Once the gravitational wave arrives, the comoving coordinates can still serve as coordinates, that is, as a means of assigning a tuple $x, y$ as a unique identifier to each point in our transverse plane. But coordinate differences will no longer correspond to physical distances. Instead, as we saw in Sec.~\ref{sec:basic-gw}, the gravitational wave stretches distances in the $x$ direction by the factor $a_x(t)$, so a comoving coordinate interval $\Dd x$ will correspond to a physical distance $\Dd s=a_x(t)\cdot\Dd x$.

We had already associated a comoving clock with each of the free-floating particles, and used those clocks to define our time coordinate.  Let us go one step further and for each of the particles, imagine a comoving observer, who can perform basic (local) measurements. Since the particle, and hence the comoving observer, are in free fall, Einstein's equivalence principle applies: For a light-signal passing by at time $t$, such an observer will find that, as measured by their own clock (which is the same as the local time given by our time coordinate $t$) and their local meter stick, the signal travels at the usual constant speed $c$.

This fact allows us to find the trajectory $x(t)$ of our light pulse, in terms of the time coordinate $t$ and comoving space coordinate $x$ we had defined: Let $\Dd t$ be the infinitesimal time interval, as measured by the local comoving clock, that it takes for the light to traverse the coordinate interval $\Dd x$. The physical distance corresponding to $\Dd x$, such as our comoving observer will measure with their local meter stick, is $\Dd s = a_x(t)\cdot \Dd x$. The equivalence principle then tells us that 
\be
\frac{\Dd s}{\Dd t} = a_x(t)\frac{\Dd x}{\Dd t} = c \;\;\Rightarrow\;\;\; c\frac{\Dd t}{a_x(t)} = \Dd x.
\ee
We can readily integrate this to obtain
\be
x-x_i = c\int\limits_{t_i}^t \frac{\Dd t'}{a_x(t')},
\label{eq:basicLightPropagationPre}
\ee
where $t_{i}$ is the time the light signal leaves its initial location at comoving coordinate value $x_i$, and $t$ the time it arrives at the comoving coordinate value $x$. For light propagating in the $y$ direction, the same reasoning applies, but with $a_y$ instead of $a_x$.

\subsection{Doppler formula}
\label{sec:DopplerSectionGW}

Next, consider the following set-up: Within our family of free-floating particles, we select two particles $a$ and $b$, both located on the $x$ axis, at locations $x_a$ and $x_b$, respectively. Then, we send two light pulses in quick succession from $a$ to $b$. Designate the time at which the first light pulse is emitted by the particle $a$ as $t_a$, and the time of emission of the second light pulse as $t_a+\delta t_a$. Conversely, denote the time the first pulse arrives at the particle $b$ as $t_b$, and the arrival time of the second pulse as $t_b+\delta t_b$. From (\ref{eq:basicLightPropagationPre}), we know that
\be
x_b-x_a = c\int\limits_{t_a}^{{t_b}} \frac{\Dd t'}{a_x(t')} = c\int\limits_{t_a+\delta t_a}^{t_b+\delta t_b} \frac{\Dd t'}{a_x(t')},
\ee
since both light pulses start out at the $x$ coordinate value $x_a$ and arrive at $x_b$. It follows that
\be
\int\limits_{{t_a+\delta t_a}}^{{t_b+\delta t_b}} \frac{\Dd t'}{a_x(t')} - \int\limits_{{t_a}}^{{t_b}} \frac{\Dd t'}{a_x(t')}  = 0.
\label{eq:LightPropagationDifference}
\ee
The limits of the first integral can be rewritten as 
\be
\int\limits_{{t_a+\delta t_a}}^{{t_b+\delta t_b}}=\int\limits_{{t_a+\delta t_a}}^{{t_a}}+\int\limits_{{t_a}}^{{t_b}}+\int\limits_{{t_b}}^{{t_b+\delta t_b}}=\int\limits_{{t_a}}^{{t_b}}-\int\limits_{{t_a}}^{{t_a+\delta t_a}}+\int\limits^{{t_b+\delta t_b}}_{{t_b}}.
\ee
In all the situation we will consider in the following, $\delta t_a$ and $\delta t_b$ will be short compared with the time scale for any change of $a_x(t)$, which means we can use the mean value theorem for definite integrals,
\be
\int\limits^{{t_a+\delta t_a}}_{{t_a}}\frac{\Dd t'}{a_x(t')} \approx \frac{\delta{t_a}}{a_x({t_a})}
\ee
and similarly for the $t_b$ integral. Putting everything together, (\ref{eq:LightPropagationDifference}) is transformed to
\be
\frac{\delta t_a}{a_x(t_a)} ={\frac{\delta t_b}{a_x(t_b)}.}
\label{eq:deltaTShift}
\ee
Evidently, the distances between successive light pulses change in the same way as the distances between our free-floating particles: in proportion to $a_x(t)$.

Now, instead of two successive light pulses, consider a sinusoidal light wave propagating from particle $a$ to particle $b$. We can choose $\delta t_a$ to be the time interval between the emission of one wave crest and the next, which means that $\delta t_b$ is the time interval between the arrivals of those two wave crests at particle $b$. In terms of the wavelength $\lambda_a$ as the wave is emitted at $a$ and the wavelength $\lambda_b$ as it arrives at $b$, we then have $\lambda_a=c\cdot \delta t_a$ and $\lambda_b=c\cdot \delta t_b$, respectively, and (\ref{eq:deltaTShift}) becomes
\be
\frac{\lambda_a}{a_x(t_a)} = {\frac{\lambda_b}{a_x(t_b)}.}
\label{eq:DopplerShiftFormula}
\ee
In terms of the gravitational strain $h$,
\be
\frac{\lambda_b}{\lambda_a} = \frac{1+\frac12 h(t_b)}{1+\frac12 h(t_a)}\approx 1 +\frac12\left[h(t_b)-h(t_a)\right].
\label{eq:DopplerFormulaWavelength}
\ee
The relative wavelength shift between emission and arrival is usually called $z$, which in astronomy goes by the name ``redshift'' (blueshifts are called ``negative redshifts'' in this context), and is
\be
z\equiv \frac{\lambda_b-\lambda_a}{\lambda_a}.
\ee
From (\ref{eq:DopplerFormulaWavelength}), we have
\be
\frac{\Delta\lambda}{\lambda}\equiv z =\frac12\left[h({t_b})-h({t_a})\right].
\label{eq:DopplerFormulaZ}
\ee
This is a key result for the ``Doppler shift'' induced by a gravitational wave: $z$ depends both on the gravitational wave's state (that is, its amplitude, orientation and phase) at the time of emission of the light and on its state at the time of the light's arrival.

We can also express the change in terms of the period $P$ of our simple sinusoidal light wave. $P$ is the time interval between the arrival of two consecutive maxima, and thus we have $P=\lambda/c$. The relative change of that period due to a gravitational wave is
\be
\frac{\Delta P}{P} = z = \frac12\left[h({t_b})-h({t_a})\right].
\label{eq:DopplerFormulaP}
\ee
The same formula applies to any periodic signal, and we will revisit it when we consider pulsar timing in section~\ref{sec:PTA}. For the same sinusoidal light wave, its frequency $f=1/P$ and wavelength $\lambda$ are linked by $f\cdot\lambda = c$, so
\be
\frac{\Delta f}{f} = -\frac{\Delta\lambda}{\lambda}=-z= \frac12\left[h({t_a})-h({t_b})\right].
\label{eq:DopplerFormulaF}
\ee
This version will become important as we consider the (as yet unrealised) detection of gravitational waves using space probes in section~\ref{sec:transponders}.

\subsection{Phase formula}
\label{sec:phaseFormula}

In addition to the various kinds of Doppler effect described in Sec.~\ref{sec:DopplerSectionGW}, we can use the light propagation equation (\ref{eq:basicLightPropagationPre}) to deduce phase information for sinusoidal light waves. Later on, we will want to describe interferometers such as LIGO, we will model elements such as light sources, beam splitters, mirrors and detectors, as free-floating particles, whose distances from each other change under the influence of a passing gravitational wave. We will restrict our analysis to the simplest case, when all these elements are in the same plane, transverse to the direction of propagation of the gravitational wave, and where light propagates as a sinus wave either in the $x$ or in the $y$ direction.

For simple sine waves like our light waves, time $t$ and phase $\varphi(t)$ are related as $\varphi(t)=2\pi f\:t+\varphi_0$, with $f$ the frequency of the wave. Thus, to relate phases at different locations --- say, the phase at the light source and at a distant detector --- we will need to be able to tell how much time our light requires to travel from a starting point to an end point.

For travel along the $x$ direction, this amounts to asking: At what time $t$ will a light signal that has left the starting point at $x_0$ at a time $t_0$ reach the coordinate value $x$? As it stands, eq.~(\ref{eq:basicLightPropagationPre}) provides an answer to the converse question, namely the coordinate value $x$ the light has reached at the time $t$, when previously at time $t_i$ it was at $x_i$:
\be
x = {x_i}\pm c\int\limits_{{t_i}}^t\frac{\Dd t'}{a(t')} = {x_i}\pm c\int\limits_{{t_i}}^t\frac{\Dd t'}{{1+\frac12 h(t')}}
\label{eq:basicLightPropagation}
\ee
where the signs correspond to propagation in the positive and negative $x$ direction, respectively. Our next task will be to find an approximate solution for $t$ in terms of $x$.

Discarding terms higher than linear order in $h$, (\ref{eq:basicLightPropagation}) can be rewritten as
\be
x = {x_i}\pm  c(t-{t_i}) \mp \frac{c}{2}\int\limits_{{t_i}}^t h(t')\:\Dd t'.
\label{eq:basicLightPropagationCoord}
\ee
For a better understanding of this equation, consider the rate of change
\be
\frac{\Dd x}{\Dd t} = 
\pm c \left(
1-\frac12 h(t)
\right).
\label{eq:CoordinateSpeedOfLight}
\ee
This is the  ``coordinate speed of light,'' that is, the rate at which the $x$ location of a light pulse or light wave, as expressed in our comoving coordinates, changes with $t$. In the absence of a gravitational wave (that is, for $h(t)=0$), both our $x$ coordinate and our $t$ coordinate revert to the usual coordinates of an inertial system in special relativity, so light moves at the speed $c$. The presence of the $h(t)$ term introduces a small variation of the coordinate speed over time. Note that, since $|h|\ll 1$, the overall sign of this expression does not change: Light moving either in the positive or negative $x$ direction does not change direction as the gravitational wave passes.

As a first step towards solving (\ref{eq:basicLightPropagationCoord}) for $t$, re-write 
(\ref{eq:basicLightPropagationCoord}) as
\be
t={t_i}\pm \frac{x-{x_i}}{c} +\frac{1}{2}\int\limits_{{t_i}}^t h(t')\:\Dd t'.
\label{eq:tOfXPre}
\ee
From (\ref{eq:CoordinateSpeedOfLight}) we know that the function $x(t)$ is strictly monotonic, hence invertible. Retaining only terms that are first-order in $h(t)$ in expression (\ref{eq:tOfXPre}), we replace the integration limit by $t-t_i \pm \frac{x-x_i}{c}$, and make a change of variables from $t'$ to $x'$ using $\Dd t' = \pm \frac{\Dd x'}{c}$, to obtain
\bea
t &=&{ t_i \pm \frac{x-x_i}{c} +\frac{1}{2} \int\limits_{t_i}^{t_i \pm \frac{x-x_i}{c}} h(t') \,\Dd t'}\\[0.5em]
&=& {t_i}\pm \frac{x-{x_i}}{c} \pm \frac{1}{2c}\int\limits_{{x_i}}^x h\!\left(t_i\pm \frac{x'-{x_i}}{c}\right)\:\Dd x'.
\label{eq:PhaseFormula}
\eea
Let us call this the {\em phase formula} for light propagation influenced by a gravitational waves. It is the counterpart to the Doppler formula in its various guises (\ref{eq:DopplerFormulaZ}), \ref{eq:DopplerFormulaP}) and (\ref{eq:DopplerFormulaF}). We can get from one to the other by integrating or differentiating. Which formula best describes how gravitational waves are detected will depend on the setup. Doppler and phase formula are two sides of the same physical coin.

\subsection{Relation to cosmology}

From a pedagogical point of view, the parallels between light propagation under the influence of a gravitational wave and in an expanding universe are worthy of note. Recall that, in the standard description of an expanding universe, distances between distant galaxies are proportional to a time-dependent cosmic scale factor $a(t)$. The pattern of motion this imposes on galaxies is called the Hubble flow. We can introduce a space-filling family of point-like, idealized galaxies, ``fundamental particles of cosmology'' whose motion follows the Hubble flow exactly. The usual cosmological time coordinate can then be defined with reference to that family: At each point in space, cosmic time is time as measured on the comoving clock of the local Hubble-flow galaxy. Global synchronisation for those comoving clocks is provided by making use of the homogeneity of the universe in these cosmological models: the comoving clocks are synchronized so that at any given moment in cosmic time, all observers in Hubble-flow galaxies will measure the same value for the mean density of the universe around them.\cite{RindlerRelativity1} 

Comoving coordinates in an expanding universe can be introduced using that same family of idealized Hubble-flow galaxies. Consider a snapshot of the universe at some fixed cosmic time $t_0$. Spatial relations within that snapshot can be described using suitable coordinates; in the simplest case, that of a spatially flat universe, we could assign to each galaxy standard Cartesian coordinates. The only change in spatial relations relative to that $t=t_0$ snapshot will be one of overall scale. That allows us assign each galaxy its snapshot coordinate at $t=t_0$ as a permanent, ``comoving'' coordinate, valid at arbitrary times $t$. The only drawback is that for $t\ne t_0$, the distances associated with those spatial coordinates do not correspond to physical distances. That is easily remedied, though: We can go from comoving distances (which, after all, correspond to physical distances at $t=t_0$) to physical distances at any time $t$, simply by rescaling with $a(t)/a(t_0)$.\cite{RindlerRelativity2} 

The parallels with the description of gravitational waves in Sec.~\ref{sec:basic-gw} should be evident, and the simplified calculation for light propagation based on the equivalence principle works in either case: Our family of free-floating test particles correspond to Hubble-flow galaxies; the time coordinate we defined corresponds to cosmic time, and both situations can be described using comoving coordinates. The direction-dependent scale factors $a_x(t)$ and $a_y(t)$ correspond to a single, universal scale factor $a(t)$ in cosmology, which governs cosmic expansion. The Doppler formula (\ref{eq:DopplerShiftFormula}) yields the standard cosmological redshift, and the analogue of (\ref{eq:basicLightPropagationPre}) describes light propagation in an expanding cosmos. In the end, both seemingly different situations are governed by the same physics, rooted in general relativity.

\section{Gravitational wave detection with transponders and space probes}
\label{sec:transponders}

\begin{figure}
\includegraphics[width=\linewidth]{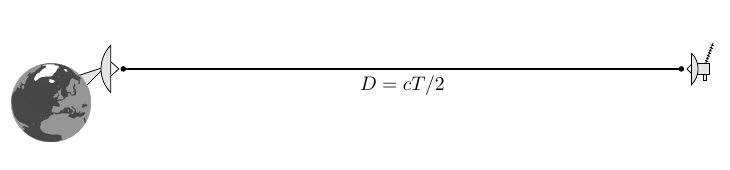}
\caption{Basic setup for detecting a gravitational wave using a spacecraft with transponder
\label{fig:transponder-setup}
}
\end{figure}

The basic setup for our first example is shown in Fig.~\ref{fig:transponder-setup}: an antenna on Earth (left) sends a radio signal with frequency $f$ to a distant space probe (right); the probe's transponder immediately sends the radio signal back. Call the Earth-to-probe distance $D=cT/2,$ with $T$ the total two-way travel time. 
As in the previous sections, we only consider plane gravitational waves propagating orthogonal to the radio signal. For the Doppler shift due to the gravitational wave, it doesn't matter whether the radio signal is travelling in the positive or in the negative $x$ direction. The total effect will only depend on the time the signal was emitted and on its total travel time; while it is travelling, the signal will be Doppler-shifted regardless of whether it is moving to the left or to the right. Back on Earth, we monitor the frequency of the returning radio signal. By the Doppler formula (\ref{eq:DopplerFormulaF}), the relative frequency shift at reception time $t$ is
\be
\frac{\Delta f}{f} =  \frac12\left[h(t-T)-h(t)\right].
\label{eq:TransponderDeltaF}
\ee
Measuring the difference $h(t-T)-h(t)$ instead of $h(t)$ directly requires extra analysis, since there are two contributions to the Doppler shift, from $h(t)$ and $h(t-T)$. But there is an interesting type of signal where the analysis is straightforward: a transient gravitational wave signal, like the chirp signal from two merging black holes. Assume that this transient gravitational wave begins to make its influence felt in the transverse plane where Earth, the space probe and the radio signal are located, at time $t_a$, and that the gravitational-wave-induced changes are complete by the time $t_b$, more specifically: that $h(t)=const.$ (although not necessarily zero!\cite{GWMemory}) for $t>t_b$. 

In the Doppler shift formula (\ref{eq:TransponderDeltaF}), there are two ``copies'' of that signal: first $h(t)$ and later on, delayed by a time interval $T$, the term $h(t-T)$. If the delay $T$ is large enough, $t_b-t_a<T$, then the $h(t)$ copy of the signal will have run its course before the term $h(t-T)$ has even begun to diverge from zero. In that case, we see two copies of the transient gravitational wave signal cleanly separated from each other: first the transient signal in $h(t)$, then a pause where nothing happens, and then a repeat of the signal via the term $h(t-T)$.

The basic principles for this kind of detection were worked out half a century ago.\cite{Kaufmann1970,Estabrook1975} 
Since then, transponder measurements in search of gravitational waves have been made using a number of interplanetary spacecraft, notably the Cassini probe. So far, those measurements have not been sensitive enough to detect gravitational waves. But whenever gravitational waves are searched for at a given sensitivity, but not detected, the results provide an upper limit for the strength of gravitational wave signals in the frequency range that was covered by the measurements.\cite{Armstrong2006} A positive result, that is, a direct detection of gravitational waves using this method, is likely to take another decade. Planned space missions to the ice giants, with a travel time of about 10 years, would provide an opportunity for these kinds of measurement.\cite{Soyuer2021} Detections should be feasible for gravitational-wave frequencies between about $10^{-5}$ and $1\;\mbox{Hz}$. Between the start of such a mission around 2030 and its arrival at Uranus or Neptune about 10 years later, transponder measurements might detect merging supermassive black holes, a stellar-size black hole merging with a supermassive black hole (extreme mass ratio inspiral, EMRI), or merging stellar-mass black holes. 

Substituting $h(t)=h_0\cdot \sin(2\pi f_{\mathrm{gw}}\:t),$ we can estimate the sensitivity of such a transponder-detector to monochromatic gravitational waves with various frequencies $f_{\mathrm{gw}}$. By the addition theorem for two sine functions, the relative frequency shift (\ref{eq:TransponderDeltaF}) at reception time $t$ is 
\bea
\nonumber \frac{\Delta f}{f} &=&\frac{h_0}{2} \Bigl[\sin(2\pi f_{\mathrm{gw}}\:[t-T])-\sin(2\pi f_{\mathrm{gw}}\:t)\Bigr]\\[0.5em]
&=& -h_0\cos\bigl(2\pi f_{\mathrm{gw}}[t-T/2]\bigr)\:\underbrace{\sin(\pi f_{\mathrm{gw}} T)}_{h_c},
\label{eq:transponderSensitivity}
\eea
This is a sine wave with an  amplitude modulation $h_0\cdot h_c$. When $h_c$ it is zero, our transponder is not able to detect the gravitational wave at all. This is because what we measure is the difference $h(t-T)-h(t)$: If the gravitational wave period is such that $t$ and $t-T$ correspond to identical phases of the wave, the difference is identically zero. To gravitational waves with frequency $f_{\mathrm{gw}} = n/T,\: n\in\mathbb{N}$, our setup is completely insensitive!

\begin{figure}
\vspace*{0.5em}
\includegraphics[width=0.9\linewidth]{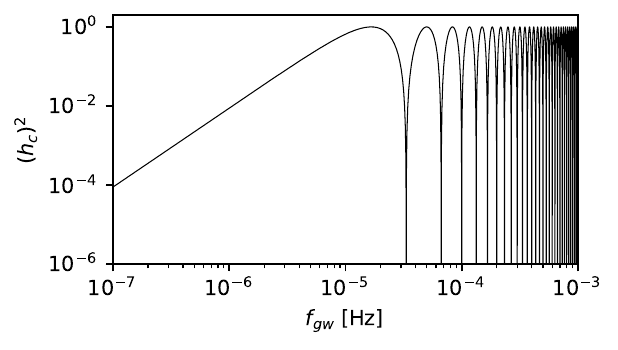}\\[-1em]
\caption{Sensitivity of a transponder set-up for a probe near Neptune, $T=30~000\:\mbox{s}$, as a function of gravitational-wave frequency
\label{fig:logarithmicHC}
}
\end{figure}

Fig.~\ref{fig:logarithmicHC} shows a logarithmic plot of the dependence of $(h_c)^2$ on $f_{\mathrm{gw}}$.
The $T$ was chosen so as to roughly correspond to a probe near Neptune, at about $30\:\mbox{au}$, with a round-trip time of roughly 8 hours. This is our first encounter with a general feature of gravitational wave detectors: linear growth due to $\sin(x)\approx x$ in (\ref{eq:transponderSensitivity}) as long as the detector time scale is small compared to the gravitational wave period, followed by periodic sensitivity drops where the gravitational wave period is an integer multiple of the detector time scale.
Sensitivity curves in the literature, e.g. Fig.~2  in ref.~\onlinecite{Soyuer2021}, will look similar, but usually not identical --- they average over different orientations of the detector relative to the gravitational wave, whereas we only consider the orthogonal case. 

\begin{figure}
\includegraphics[width=\linewidth]{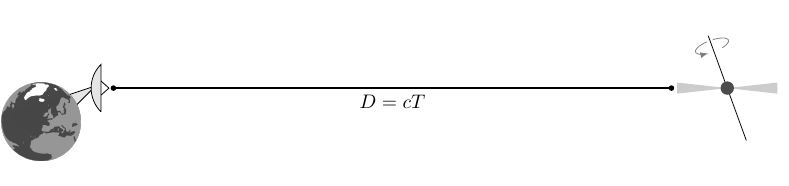}
\caption{Basic setup for the detection of a regular pulsar signal. The directional radio signals from the pulsars are shown as cones
\label{fig:pulsar-setup}
}
\end{figure}

\section{Pulsar timing arrays}
\label{sec:PTA}
Next, consider the case where our radio signal is not artificial, but the highly regular series of pulses reaching us from a distant pulsar. The situation is sketched in Fig.~\ref{fig:pulsar-setup}.\cite{McGrath2021} The pulses we receive on Earth are created through a light-house effect: The pulsar emits intense radio waves in two opposite directions. Those directions are not aligned with the pulsar's rotational axis, and thus trace out two cones as the pulsar rotates. If the pulsar is aligned just right, we receive a radio pulse every time the pulsar's beam brushes across Earth.

For $T$ the (one-way) travel time of the signal from the pulsar to Earth, the distance is $D=cT$. Pulse times-of-arrival (TOA) are recorded by radio telescopes utilising precise atomic clocks. Those TOA measurements require summing up a considerable number of consecutive pulses in a coherent way, and fitting a suitable template for the shape of the pulse to the result. Uncertainties in this fitting procedure are the main source of uncertainty for TOA measurements. The TOA sequence is then decomposed into a regular part, corresponding to the period $P$, the correction due to (constant) period drift $\dot{P}$, and a time-dependent correction $\Delta P$ that changes the time interval between the arrival of each pulse and its successor. In the following simplified toy model, we will instead pretend that our radio telescope is determining times-of-arrival for single, separate pulses, that we have already determined $P$ in the absence of gravitational waves, and that $\dot{P}=0$, so that we can now set out directly to measure the influence of the gravitational wave by considering the $\Delta P$.

The most stable known pulsars are {\em millisecond pulsars}, with rotation periods $P$ on the scale of milliseconds. Those pulsars have been spun up to their high rotational speed by accreting material from a companion star, and $P$ is three to four orders of magnitude more stable than for the much more common younger pulsars with rotation periods of about a second.\cite{Verbiest2009} From the Doppler formula (\ref{eq:DopplerFormulaP}), it follows that the period shift for a pulsar signal arriving at time $t$ that has travelled to us along the $x$ axis is
\be
\frac{\Delta P}{P}  = \frac12\left[h(t)-h(t-T)\right].
\label{eq:DopplerFormulaPulsar}
\ee
What we can measure is once more the difference of two terms: the state of the gravitational wave $h(t)$ as the pulse arrives at Earth and its state $h(t-T)$ at the time the pulse left the pulsar. 

For the moment, let us ignore $h(t-T)$, focus on $h(t)$, and consider the accuracy needed for detecting $\Delta P$. The current PTA measurements detect a ``stochastic background,''  the combination of many unresolved gravitational wave events in various locations in the cosmos. In our toy model, we replace this background by a continuous sine wave with maximum amplitude $h/2\sim 10^{-14}$. For a millisecond pulsar, say $P=5\;\mbox{ms}$, this would mean $\Delta P=5\cdot {10^{-8}}\;\mbox{ns}$. Even in a fictitious world where we could determine pulse times-of-arrival with an accuracy of $1\:\mbox{ns}$, this order of magnitude would be impossible to detect.

But there is hope. Let us, just for the moment, consider $\Delta P>0$ as constant over the observation time --- corresponding to an extremely low-frequency gravitational wave. Using our clock to document pulse arrival time, we find the second pulse is $\Delta P$ later than expected. The third pulse will be late by $2\cdot \Delta P$, and the $(n+1)$th pulse by $n\cdot\Delta P$. Even where $\Delta P$ itself is undetectable, a shift $n\cdot\Delta P$ with large $n$ need not be. In our example, after 500 days (about $4\cdot 10^{7}$ seconds, or $n=8\cdot 10^9$ periods), the pulse is late by $400\:\mbox{ns}$, a shift that current TOA measurement techniques, whose accuracy can be better than $100\:\mbox{ns}$, would be able to detect.

\begin{figure}
\includegraphics[width=\linewidth]{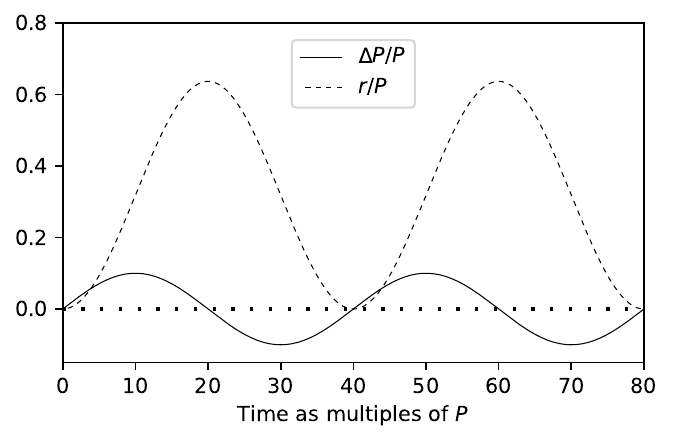}
\caption{Timing residuals for a sine gravitational wave with $1/f_{\mathrm{gw}}=40\:P$ and the unrealistically large amplitude $h_0=0.1$
\label{fig:timing-residual-plot}
}
\end{figure}

Returning to non-constant $\Delta P$, we define the {\em timing residual} $r$ as the cumulative shift at time $t$. For $P$ sufficiently small over our measuring interval from $t_0$ to $t$, we can write the $N$ summed-up individual period shifts as an integral,
\be
r = \sum_{i=1}^N\Delta P(t_i) = \sum_{i=1}^N \frac12 h(t_i)\cdot P
\approx \frac12 \int\limits_{t_0}^t h(t')\Dd t'.
\ee
To see how this works, consider the simple example of a monochromatic (sine) gravitational wave with frequency $f_{\mathrm{gw}}$ with amplitude $h_0=0.1$ and $1/f_{\mathrm{gw}}=40\:P$. For such a signal with $h(t)=h_0\sin(2\pi f\cdot t)$,
\bea
{r}&{\approx}& {\frac{h_0}{2}\int\limits_{t_0}^t \sin(2\pi f\cdot t')\Dd t'}\\[0.2em]
&{=}& {\frac{h_0}{4\pi f}[\cos(2\pi f\cdot t_0)-\cos(2\pi f\cdot t)],}
\eea
demonstrating how the detection is more sensitive for smaller gravitational-wave frequencies $f$. Fig.~\ref{fig:timing-residual-plot} shows how this timing residual amplifies the effect of the individual period shifts up to a maximum that consists of all the positive shifts $\Delta P$ of the sine wave adding up.

This leaves us with one remaining problem: We neglected the delayed term $h(t-T)$ in (\ref{eq:DopplerFormulaPulsar}). We can solve the problem by considering not a single pulsar, but a larger set of pulsars: a {\em pulsar timing array} (PTA). These pulsars will naturally be at different locations in space. In our one-dimensional toy model, the pulsars would be at different distances in the $x$ direction, with different pulse travel times $T_i$. Adding up their various timing residuals, we obtain
\bea
\nonumber
r_{\mathrm{PTA}} &=& \sum_{j=1}^{N} \frac12\int\limits_{t_0}^t \left[
h(t') - h(t' - T_j)
\right]\Dd t'\\[0.2em]
&=& \frac{N}{2}\int\limits_{t_0}^t 
h(t') \Dd t' -  \sum_{j=1}^{N}  \frac12\int\limits_{t_0}^t  h(t'-T_j)\Dd t'. 
\label{eq:residualsAddedUp}
\eea
The first term of the sum, the ``Earth term'' $h(t)$ describing the phase of the gravitational wave at the time the pulses simultaneously arrive on Earth, has received a boost in our simplified situation: it is amplified by the number $N$ of pulsars in the set. In the remainder, each part of the sum is a ``pulsar term,'' describing the state of the gravitational wave at the event when the pulse left a particular pulsar. The pulsar term contributions to the sum will have different phases --- some will be positive, some negative. Those terms average out, so the transition from a single pulsar to an array of pulsars has indeed solved our problem.

For a realistic three-dimensional pulsar timing array, the analysis is more complicated, but an important part of what is going on is covered by the toy model: There, too, documenting timing residuals over a sufficiently long time will boost the signal. The more complicated geometry, with the pulsars distributed all over the sky, means that the toy model's simple adding-up of residuals  (\ref{eq:residualsAddedUp}) is not sufficient. We require an additional step: First, pairs of pulsar signals are correlated. Then, those correlations are averaged, which has the same effect as in (\ref{eq:residualsAddedUp}) of the pulsar terms averaging out. The usual averaging-out procedure requires the fact that the real gravitational-wave background is not a sine wave from a well-defined direction, as in our toy model, but a stochastic mix of signals reaching us from all possible directions in the sky.\cite{Hellings1983,Romano2024} This kind of analysis is the basis for the June 2023 announcements of various PTA collaborations.\cite{PulsarTiming2023a,PulsarTiming2023b,PulsarTiming2023c,PulsarTiming2023d} If instead of a stochastic background signal, the array were to look at a monochromatic signal from a localized source, then at least for nearby sources, a more complicated PTA analysis could make use of both the Earth term and the pulsar term to reconstruct $h(t)$.\cite{Lee2011} For a transient instead of a monochromatic signal, the same reasoning would apply as for the transponder method in Sec.~\ref{sec:transponders}: given a sufficiently brief transient signal, we would first detect the Earth term, and then after a pause (which, depending on the pulsar distance, would likely defy the time scale of any human research project) the pulsar term.

\section{LIGO, LISA et al.}
\label{sec:interferometricDetectors} 
\subsection{Michelson interferometer basics}
\label{sec:michelson}
\begin{figure}
\includegraphics[width=0.8\linewidth]{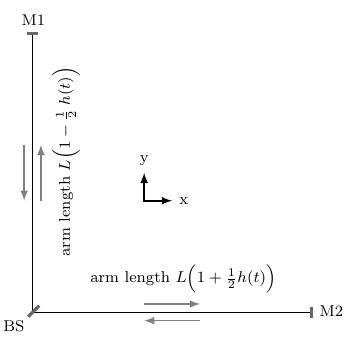}
\caption{Basic setup for a Michelson interferometer influenced by a gravitational wave
\label{fig:michelson-gw}
}
\end{figure}

The basic setup of an interferometric detector like LIGO or Virgo is that of a Michelson interferometer as in Fig.~\ref{fig:michelson-gw}. %
Monochromatic light from a light source S propagates to a beam splitter BS; half of the light takes a round trip via the mirror M1, the other via the mirror M2. On returning to BS, the light portions are combined coherently, and some portion of combined light reaches the photodetector D. In a detector like LIGO, the optical elements are suspended so that we can treat their motion, at least when it comes to the distances between BS and M1/M2, as free-fall motion. That is fortunate: For our simple linearly-polarized gravitational wave propagating orthogonal to the detector plane, we can use the formalism we developed in sections \ref{sec:basic-gw} and \ref{sec:GW-light}. 

Let BS and M2 be separated in the $x$ direction, BS and M1 in the $y$ direction. 
When the two portions of light leave the beam splitter BS, they have the same phase. For concreteness, let us describe the amplitude of the light wave (e.g.\ the electric field in z direction) emanating from BS into each arm by a sine function $A\cdot \sin(2\pi f\cdot t)$, with $f$ the frequency of the laser light.

For the remainder of this section, assume a constant length offset, with
$\overline{\mathrm{BS}\; \mathrm{M1}}=L$ and $\overline{\mathrm{BS}\; \mathrm{M2}}=L+\Delta L$. Ignoring discrete phase jumps associated with reflection,
the light arriving {\em back} at BS which makes its way to the detector D is then given by
\be
\frac{A}{2}\big\{
\sin\left(2\pi f [t-2L/c ]\right)
+\sin\left(2\pi f[t-2(L+\Delta L)/c]\right)
\big\}
\ee
--- a coherent superposition that takes into account the different round-trip times for light in the two arms. With the sum formula for sine functions, this is
\be
A\sin\left(2\pi f[t-2L/c-\Delta L/c]\right)\cdot\cos\left(2\pi \Delta L/\lambda\right),
\label{eq:michelsonFormula}
\ee
where we have used $f\cdot\lambda=c$ for the light's frequency $f$ and wavelength $\lambda$. The first factor in this product is the same as the original form of our light wave, except for a constant phase shift. A photodetector integrates light power over numerous periods of oscillation, so this term does not influence our measurement at all. The second factor modulates the light signal as a whole, and its square will determine how much power we observe. If $\Delta L=0$ were the default state and $\Delta L\ne 0$ what we are trying to detect, a cosine, or the square of a cosine once we look at the brightness of the light, has two disadvantages. One is that $\Delta L=0$ corresponds to a maximum, so detecting a gravitational wave would amount to detecting a small brightness change of  already rather bright light. It is easier to detect a small change in the dark, and thus advantageous to adjust the detector so that $\Delta L=0$ corresponds to complete dark or near-dark.\cite{DarknessRemark} Shifting Mirror 2 by $-\lambda/4$ transforms the cosine into a sine, 
\be
\cos\left(2\pi [\Delta L-\lambda/4]/\lambda\right) 
=\sin\left(2\pi\Delta L/\lambda \right).
\ee
Done? No, since the new sine-square function for the brightness still has an extremum at $\Delta L=0$, which means its response to $\Delta L\ne 0$ is a second-order effect! An additional small length change in one arm, corresponding to a phase shift $\Delta\phi$, helps. With the shifted argument
\be
\chi \equiv 2\pi \frac{\Delta L}{\lambda}+\Delta\phi,
\ee
we have 
\be
\sin^2\chi \approx \sin^2\left( \Delta\phi\right) + \sin\left( \Delta\phi \right) \cos\left( \Delta\phi \right) 4\pi \frac{\Delta L}{\lambda}.
\ee
Now the change in the power signal we measure is directly proportional to the length change $\Delta L$. We will take this as the operating principle of our interferometric detector.\cite{ComplexComment} Our take-away is: Whenever we end up with our light signal modulated by a cosine as in (\ref{eq:michelsonFormula}), suitable constant phase shifts can make it so that the argument of that cosine is proportional to the gravitational wave signal we detect.

\subsection{Interferometric GW detectors}
\label{sec:interferometric}

Now, we extend our analysis to length changes caused by gravitational waves. We again consider the layout shown in Fig.~\ref{fig:michelson-gw}, but this time with gravitational-wave-induced length changes given by (\ref{eq:SimpleFormAandH}) and (\ref{eq:SimpleFormAandHY}), respectively.

Just as in the previous section, in order to determine the action of the gravitational wave on our interferometer, we need to determine when light that is returning to the beamsplitter {\em now}, at time $t$, originally started out at the beamsplitter --- which will tell us the relative phase of the two portions of light arriving at the beamsplitter simultaneously, at the present time (and, in part, going on to the photodetector). With the help of (\ref{eq:PhaseFormula}), we can express the beamsplitter start time $t_i$ in terms of the arrival time. An advantage of the simplicity of our situation, where light travels orthogonally to the propagation of the gravitational wave, is that we do not need to worried about the travel back and forth; for calculating the travel time, the result for travelling from $0$ to $L$ and back is the same as for travelling in a single direction from $0$ to $2L$. All in all, for light travelling from BS to M2 and back along the $x$ direction, we have 
\be
t_{i} =  t-\frac{2L}{c} -\frac{1}{2c}\int\limits_0^{2L}h\left(t+\frac{x'-2L}{c}\right)\Dd x' .  
\ee
The corresponding formula for the $y$ arm is obtained via $h\mapsto -h$. Coherent superposition amounts to adding up the two sine contributions from the two arms, as
\begin{widetext}
\begin{eqnarray}
\nonumber \!\!\!\!&&\!\!\!\!\!\!\!\!\!\!\!\!\!\!\sin\left(2\pi f\left[t-\frac{2L}{c} -\frac{1}{2c}\int\limits_0^{2L}h\left(t+\frac{x'-2L}{c}\right)\Dd x'\right]\right) + \sin\left(2\pi f\left[t-\frac{2L}{c} +\frac{1}{2c}\int\limits_0^{2L}h\left(t+\frac{x'-2L}{c}\right)\Dd x'\right]\right)\\[0.5em]
\!\!\!\!&=&2\sin\left(2\pi f\left[t-\frac{2L}{c}\right]\right)\cos\left(\frac{\pi f}{c}\int\limits_0^{2L}h\left(t+\frac{x'-2L}{c}\right)\Dd x'\right).
\end{eqnarray}
\end{widetext}
This is the same kind of cosine term as in (\ref{eq:michelsonFormula}), and using the same constant length offsets as discussed in sec.~VI.A, our detector response will be proportional to the argument of the cosine. For a sine-shaped gravitational wave with $h(t)=h_0\cdot\sin(2\pi f_{\mathrm{gw}}\:t),$ that argument is
\be
 \frac{\pi f}{c}\int\limits_0^{2L}h\left(t+\frac{x'-2L}{c}\right)\Dd x'
 \ee
 which we can rewrite as
\begin{eqnarray}
\nonumber &\!\!\!\!\!\!\!\!\!\!&  \pi f\int\limits_{t-2L/c}^th(t')\Dd t'
=\pi f h_0\int\limits_{t-2L/c}^t\sin(2\pi f_{\mathrm{gw}}t')\Dd t'\\[0.5em]
\nonumber &\!\!\!\!\!\!\!\!\!\!=&\frac{h_0}{2}\frac{f}{f_{\mathrm{gw}}}\left[
\cos\left(\!2\pi f_{\mathrm{gw}}\!\!\left[t-\frac{2L}{c}\right]\right) - \cos(2\pi f_{\mathrm{gw}}t)
\right]\\[0.5em]
&\!\!\!\!\!\!\!\!\!\!=&
h_0\frac{f}{f_{\mathrm{gw}}}\sin\left(2\pi f_{\mathrm{gw}}\left[t-\tau_L/2\right]\right)
\sin\left(\pi f_{\mathrm{gw}}\tau_L \right).
\label{eq:ModulationGW}
\end{eqnarray}
In the last step, we have introduced the {\em detector time scale} $\tau_L\equiv 2L/c$ as a measure of light travel time inside the detector. An interesting limiting case is $f_{\mathrm{gw}}\tau_L\ll 1$: Taylor-expanding the result (\ref{eq:ModulationGW}) in $f_{\mathrm{gw}}\tau_L$, we have 
\be
\frac{2\pi}{\lambda} \: L\:h_0\: \sin(2\pi f_{\mathrm{gw}}t) = 4\pi\frac{\Delta L(t)}{\lambda},
\label{eq:ShortArmGWInfluence}
\ee
where we have substituted the wavelength $\lambda=c/f$ of the light and, on the right-hand side, 
where on the right-hand side, we have substituted the time-dependent length change in the interferometer arm $L$ caused by the gravitational sine wave. Comparison with the argument of the cosine in (\ref{eq:michelsonFormula}) shows that this is indeed the expected result for a Michelson interferometer where the difference between arm lengths is $2\Delta L(t)$, with the time-dependent length change governed by $h(t)$ as in our basic formula (\ref{eq:StrainDef}) for the physical meaning of the strain. This is known as the {\em short-arm approximation}, since re-written in terms of the gravitational wave's wavelength $\lambda_{\mathrm{gw}}=c/f_{\mathrm{gw}}$ and the arm length $L$, the Taylor approximation is valid for $L\ll \lambda_{\mathrm{gw}}.$

\begin{figure}
\vspace*{0.5em}
\includegraphics[width=\linewidth]{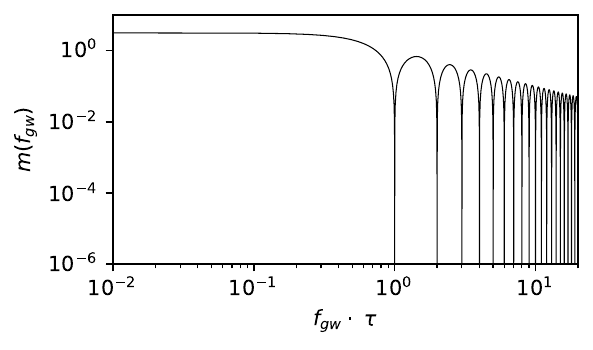}\\[-1em]
\caption{Sensitivity of an interferometric detector as per (\ref{eq:mIntDetector})
\label{fig:intSensitivity}
}
\end{figure}

Beyond that approximation, the amplitude of our detector response is governed by the factor 
\be
\frac{1}{f_{\mathrm{gw}}} \sin\left(\pi f_{\mathrm{gw}}\tau_L\right)
\label{eq:CombinedEffectInterferometer}
\ee
in (\ref{eq:ModulationGW}). Let us divide this by $\tau_L$ to make the expression dimensionless; this amounts to dividing out a linear overall ``longer arm-length is better'' factor. Discarding the overall sign, consider the dimensionless function
\be
m(f_{\mathrm{gw}}) = \left|\frac{1}{f_{\mathrm{gw}}\tau_L}\sin\left(\pi f_{\mathrm{gw}}\tau_L\right)\right|,
\label{eq:mIntDetector}
\ee
which is plotted in Fig.~\ref{fig:intSensitivity} as a function of  $f_{\mathrm{gw}}\tau_L$.
Evidently, as long as $1/f_{\mathrm{gw}}$ is considerably larger than the light travel time $\tau$, the function $m(f_{\mathrm{gw}})$ is almost constant. This is where the short-arm approximation holds. Once the two become comparable, we have periodic changes between maximum sensitivity and sensitivity zero, overlaid by a downward trend --- a ``long-arm penalty'' where, at odds with the usual ``bigger is better'' philosophy for such detectors, our detector becomes less sensitive because of its longer arms. 

So what is happening here? There are two overall effects. One is that light travel time in each arm is changed as the overall arm length changes due to the influence of the gravitational wave. The other is that the wavelength of the travelling light is affected directly while the light spends propagating in the detector. The short-arm approximation is valid when the time light spends inside the detector is short relative to the gravitational wave period. In this case, the wavelength change given by the Doppler formula (\ref{eq:DopplerFormulaZ}) is small, and can be neglected; the only remaining effect is that of the arm length difference at that particular time. In that scenario, the detector constantly uses new, ``fresh,'' unaffected light to map its differential arm length change --- and as one would expect, the result (\ref{eq:ShortArmGWInfluence}) is the same as if we were dealing with a classical Michelson interferometer, its light completely unaffected, but its arm lengths changing in the characteristic quadrupole pattern.

Beyond the short-arm approximation, both travel-time and the Doppler effects need to be accounted for, as in (\ref{eq:CombinedEffectInterferometer}). The periodic points of complete insensitivity in that formula, and Fig.~\ref{fig:intSensitivity}, can be made plausible, as well: When $\tau\cdot f_{\mathrm{gw}}=1$, light propagating once throughout the detector experiences each phase of the gravitational wave exactly once, in a completely symmetric fashion: for every stretching of the distance, the corresponding shrinking; for every Doppler stretching of the waves themselves, a corresponding blueshift. A simplified animation of what happens outside the short-arm regime can be found in ref. \onlinecite{Poessel2023}.

\section{Discussion}
\label{sec:discussion}

\begin{table*}[t]
\caption{Detector dimensions and frequency ranges
\label{tab:detectorProperties}}
\begin{ruledtabular}
\begin{tabular}{l c c c c}
detector & $L$ &  $\tau=2L/c$& $\tau_{\mathrm{gw}}=1/f_{\mathrm{gw}}$ range & $f_{\mathrm{gw}}$ range \\\hline
GEO~600 &   $2\cdot 0.6 = 1.2\:\mbox{km}$ & $8\:\mu\mbox{s}$  & $0.2\ldots 20\:\mbox{ms}$ 
& $50\:\mbox{Hz}\ldots 6\:\mbox{kHz}$ \\
LIGO & $4\:\mbox{km}$ & $27\:\mu\mbox{s}$ & $0.5\ldots 50\:\mbox{ms}$ &
$20\:\mbox{Hz}\ldots 2 \:\mbox{kHz}$ \\
LISA & $2.5\cdot 10^6\:\mbox{km}$& 17 s & $10\ldots 10^5$ s &  $0.1\:\mbox{mHz}\ldots 0.1\:\mbox{Hz}$ \\
\end{tabular}
\end{ruledtabular}
\end{table*}

For pedagogical reasons, the descriptions in the previous sections only cover certain special cases. A comprehensive description would need to include all possible gravitational wave polarisations, more general wave forms, non-planar gravitational waves, and arbitrary orientations of the detectors and of the propagating light relative to the gravitational wave. But even the simplified accounts of the various detection methods show the connections between seemingly different measurements such as pulsar timing residuals and interferometric phase comparisons. At the most fundamental level, all of the measurements that rely on gravitational waves' interaction with light and (approximately) free-fall particles can be understood in terms of the Doppler formula in its various guises (\ref{eq:DopplerFormulaZ}), (\ref{eq:DopplerFormulaP}), (\ref{eq:DopplerFormulaF}) and the phase formula (\ref{eq:PhaseFormula}), the former the differential forms of the latter.

The unified view presented here is almost at the opposite end of the spectrum from a question like ``If light waves are stretched by gravitational waves, how can we use light as a ruler to detect gravitational waves?''\cite{Saulson1997,Faraoni2007} Light-waves being doppler-shifted is what all the detection methods have in common. It is the Michelson interferometer that is the odd one out, in that changes in overall path length play an important role. The short-arm approximation, neglecting the Doppler shifts altogether, is an extreme limiting case of a more general set-up.

Almost all of the technical efforts in constructing detectors like LIGO, Virgo, KAGRA or LISA are directed towards the suppression of various sources of noise that would otherwise drown out the exceedingly weak gravitational wave signals. It is worth noting that some of the resulting design modifications do affect the applicability of the simple interferometer model from Sec.~\ref{sec:interferometric}: Consider the key dimensions of the German-British detector GEO600,\cite{GEO600Ref} the LIGO detectors,\cite{LIGORef} and LISA\cite{LISARef} summarized in Table~\ref{tab:detectorProperties}. For the ground-based detectors, GEO600 with its folded light path bringing the travel length to and from the mirrors to twice the 600 meter overall length that is part of the detector's name, ensures that the short-arm approximation is safely applicable. 

The LIGO detectors, however, are not the simple Michelson interferometers of Fig.~\ref{fig:michelson-gw}. LIGO boosts sensitivity by building each arm as a Fabry-Perot interferometer: with a probability of nearly 99\%, light heading back in the direction of the beam splitter will be reflected at an extra mirror at the inner end of the arm, taking another turn up and down that arm. Most light will spend considerably longer than $27\:\mu\mbox{s}$ in an arm. This increases laser power by a factor $G_{arm}=270$, corresponding to an average light storage time $\tau=3.6\:\mbox{ms}$, the same as for a detector with an arm length of $L_{\mathrm{eff}}=270\cdot 4\:\mbox{km}= 1080\:\mbox{km}$. 

This would seem to be well within LIGO's $\tau_{\mathrm{gw}}$ range, suggesting that the short-arm approximation is not applicable. But it's not that simple, either: Different portions of laser light will spend a different amount of time in a LIGO arm, largely averaging out the $L$-specific effects in (\ref{eq:ModulationGW}). Due to this averaging-out, the short-arm approximation works well for LIGO (and Virgo, and KAGRA) after all, and is commonly used as the basis for analysing the performance of such detectors.\cite{SaulsonBook} Discussions of the limits of the short-arm approximation and the need for corrections have a long history within the community.\citep{Cooperstock1993,Finn2009,Melissinos2010,Koop2014,Congedo2017,Lobato2021,Ruggiero2021}
 
For LISA, there is a different complication. With its three satellites forming a gigantic, approximately equilateral triangle $2.5$ million kilometers a side, LISA is clearly beyond the short-arm approximation, and the ``arm-length penalty'' imposed by (\ref{eq:ModulationGW}), with a periodic structure overlaid with overall-worsening sensitivity at higher frequencies, can be clearly seen in sensitivity plots.\cite{LISARef} 

But in practice, LISA's performance is not directly based on the phase formula (\ref{eq:PhaseFormula}). Instead, LISA's laser signals are compared to a reference laser whenever they arrive at one of the spacecraft (by superimposing both signals and measuring the beat frequency), yielding Doppler shifts (\ref{eq:DopplerFormulaF}). 
In the analysis, the Doppler shifts are integrated up, in effect making the transition from the Doppler formula (\ref{eq:DopplerFormulaF}) back to the phase formula (\ref{eq:PhaseFormula}), but with a twist: For the integration, artificial time delays are introduced, yielding the phase formula not for LISA (whose arm lengths vary by about 1\% over the course of a year) but for a virtual Michelson interferometer with equal-length arms. This process is called {\em time-delay interferometry} (TDI), and it is again related to noise suppression: In an equal-arm-length interferometer, noise due to the unavoidable jitter in the laser frequency cancels out. Suppressing laser-frequency noise by about eight magnitudes in this way is part of what will make LISA's gravitational-wave detections possible in the first place.\cite{LISA-TDI}
 
\section*{Acknowledgements}
I would like to thank Thomas M\"uller and the anonymous referees for helpful comments on earlier versions of this text, and Norbert Wex for his help with the section on pulsar timing arrays. I have no conflicts to disclose.


\begin{thebibliography}{99}
\bibitem{PulsarTiming2023a} Gabriella Agazie et al., ``The NANOGrav 15 yr Data Set: Observations and Timing of 68 Millisecond Pulsars,''  Ap. J. L. {\bf 951}, L9 (2023). \doi{10.3847/2041-8213/acda9a}
\bibitem{PulsarTiming2023b} John Antoniadis, ``The second data release from the European Pulsar Timing Array III. Search for gravitational wave signals,'' arXive e-print (2023). \doi{10.48550/arXiv.2306.16214}
\bibitem{PulsarTiming2023c} Daniel J. Reardon et al., ``Search for an Isotropic Gravitational-wave Background with the Parkes Pulsar Timing Array,'' Ap. J. L. {\bf 951}, L6 (2023). \doi{10.3847/2041-8213/acdd02} 
\bibitem{PulsarTiming2023d} Heng Xu et al., ``Searching for the Nano-Hertz Stochastic Gravitational Wave Background with the Chinese Pulsar Timing Array Data Release I.,'' Res. Astron. Astrophys {\bf 23}, 075024 (2023). \doi{10.1088/1674-4527/acdfa5}
\bibitem{ESA2024} ESA, ``Capturing the ripples of spacetime: LISA gets go-ahead'' (25 January 2024). \url{https://www.esa.int/Science_Exploration/Space_Science/Capturing_the_ripples_of_spacetime_LISA_gets_go-ahead}
\bibitem{Farr2012} Benjamin Farr, GionMatthias Schelbert, and Laura Trouille, ``Gravitational wave science in the high school classroom,'' Am. J. Phys. {\bf 80}, 898--904 (2012). \doi{10.1119/1.4738365} 
\bibitem{Choudhary2018} Rahul K. Choudhary et al., ``Can a short intervention focused on gravitational waves and quantum physics improve students’ understanding and attitude?,'' Phys. Edu. {\bf 53(6)}, 065020 (2018). \doi{10.1088/1361-6552/aae26a}
\bibitem{vanHeijningen2021} Joris van Heijningen, ``Where do gravitational waves come from and how can we detect more?'', in Magdalena Kersting and David Blair, {\em Teaching Einsteinian Physics in Schools. An Esssential Guide for Teachers in Training and Practice}, Ch.~13 (Routledge, London and New York, 2021).
\bibitem{Spetz1984} Gary W. Spetz, ``Detection of gravity waves,'' Phys. Teacher 22, 282--287 (1984). \doi{10.1119/1.2341545}
\bibitem{Ryden} E.g.\ Ch.~4 in Barbara Ryden and Bradley M. Peterson, {\em Foundations of Astrophysics} (Cambridge University Press, Cambridge, 2021).
\bibitem{LSC2015} Sec.~3 in The LIGO Scientific Collaboration (J. Aasi et al.), ``Advanced LIGO,'' Class. Quantum Grav. {\bf 32}, 074001 (2015). \doi{10.1088/0264-9381/32/7/074001} 
\bibitem{Bauchrowitz2013} J\"oran Bauchrowitz, Tobias Westphal and Roman Schnabel, ``A graphical description of optical parametric generation of squeezed states of light,'' Am. J. Phys. {\bf 81}, 767--771 (2013). \doi{10.1119/1.4819195}
\bibitem{FeynmanLectures} E.g. Ch.~29 in Richard P. Feynman, Robert B. Leighton and Matthew Sands, {\em Lectures on Physics}, Volume I (Addison-Wesley, Reading MA, 1963).
\bibitem{RindlerRelativity1} Sec.~16.2 in Wolfgang Rindler, {\em Relativity: Special, General and Cosmological} (Oxford University Press, New York, 2001).
\bibitem{RindlerRelativity2} Sec.~16.4 in Rindler, op. cit.
\bibitem{GWMemory} That, for transient signals such as black hole chirps, $h(t)$ does not return to zero afterwards, but to another (very small) constant value is known as ``gravitational wave memory.'' Cf. Kip S. Thorne, ``Gravitational-wave bursts with memory: The Christodoulou effect,'' Phys. Rev. {\bf D45}, 520--524 (1992). \doi{10.1103/PhysRevD.45.520} and references therein.
\bibitem{Kaufmann1970} William J. Kaufmann, ``Redshift fluctuations arising from gravitational waves,'' Nature {\bf 227}, 157--158 (1970). \doi{10.1038/227157a0}
\bibitem{Estabrook1975} Frank B. Estabrook and Hugo D. Wahlquist, ``Response of Doppler spacecraft tracking to gravitational radiation,'' Gen. Rel. Grav. {\bf 6}, 439--447 (1975). \doi{10.1007/BF00762449}
\bibitem{Armstrong2006} John W. Armstrong, ``Low-Frequency Gravitational Wave Searches Using Spacecraft Doppler Tracking,'' Living Rev.\ Relativity {\bf 9}, article id 1 (2006, revised 2016). \doi{10.12942/lrr-2006-1} 
\bibitem{Soyuer2021} Deniz Soyuer et al., ``Searching for gravitational waves via Doppler tracking by future missions to Uranus and Neptune,'' MNRAS {\bf 503}, L73--L79 (2021). \doi{10.1093/mnrasl/slab025}
\bibitem{McGrath2021} This section profited greatly from the clear presentation of the basics of pulsar timing arrays in Casey McGrath, {\em Gravitational wave timing residual models for pulsar timing experiments} (PhD Thesis, University of Wisconsin--Milwaukee, 2021). \doi{10.48550/arXiv.2109.07603}
\bibitem{Verbiest2009} Joris P. W. {Verbiest}, ``Timing stability of millisecond pulsars and prospects for gravitational-wave detection,'' MNRAS {\bf 400}, 951--968 (2009). \doi{10.1111/j.1365-2966.2009.15508.x} 
\bibitem{Hellings1983} Ronald Hellings and George Downs, ``Upper limits on the isotropic gravitational radiation background from pulsar timing analysis,'' ApJ {\bf 265}, L39--L42 (1983). \doi{10.1086/183954}
\bibitem{Romano2024} Joseph D. Romano and Bruce Allen, ``Answers to frequently asked questions about the pulsar timing array Hellings and Downs curve,'' Class. Quantum Grav. {\bf 41}, 175008 (2024). \doi{10.1088/1361-6382/ad4c4c}
\bibitem{Lee2011} Keija J. Lee et al., ``Gravitational wave astronomy of single sources with a pulsar timing array,'' MNRAS {\bf 414}, 3251--3264 (2011). \doi{10.1111/j.1365-2966.2011.18622.x}
\bibitem{DarknessRemark} There is another strong argument for having the default state be near dark, but it goes beyond our simplified description: For reducing the quantum noise (``photon shot noise'') mentioned in Sec.~\ref{sec:DefLight}, it proves advantageous to increase the power of the laser light. A key technique for doing this is to tune the interferometer so that the default state at the photodetector is near-dark. On its own, this would mean almost all of the laser light would leave the Michelson interferometer at the other end, in the direction of the light source. Adding a semi-transparent ``power recycling mirror'' in front of the light source, which reflects most of the returning light back into the interferometer, significantly increases the amount of laser light that is inside the detector at each moment in time.
\bibitem{ComplexComment} Reality is more complicated: The output signal is fed to actuators that actively compensate for a differential arm length change. That part of the gravitational wave that is slow enough for those actuators to keep up will be in the control signal, while the part that is too fast for the actuators will be in the signal from the near-dark photodetector.
\bibitem{Poessel2023} Markus P\"ossel, ``How an interferometric gravitational wave detector works (animation),''  {Zenodo} (2023). \doi{10.5281/zenodo.8286504}
\bibitem{Saulson1997} Peter R. Saulson, ``If light waves are stretched by gravitational waves, how can we use light as a ruler to detect gravitational waves?,'' Am. J. Phys. {\bf 65}, 501--505 (1997). \doi{10.1119/1.18578}
\bibitem{Faraoni2007} Valerio Faraoni, ``A common misconception about LIGO detectors of gravitational waves.'' In: Gen. Rel. Grav. {\bf 39}, 677--684 (2007). \doi{10.1007/s10714-007-0415-5} 
\bibitem{GEO600Ref} Benno Willke et al., ``The GEO 600 gravitational wave detector,'' Class. Quantum Grav. {\bf 19}, 1377--1387 (2002). \doi{10.1088/0264-9381/19/7/321}
\bibitem{LIGORef} Benjamin P. Abbott et al., ``Prospects for observing and localizing gravitational-wave transients with Advanced LIGO, Advanced Virgo and KAGRA,'' Liv. Rev. Relavitity {\bf 21}, 3 (2018). \doi{10.1007/s41114-018-0012-9} 
\bibitem{LISARef} LISA CDF Team, Internal Final Presentation. ESTEC, 05-05-2017. \url{https://sci.esa.int/documents/35005/36499/1567260318050-LISA_CDF_IFP_2017.pdf}
\bibitem{SaulsonBook} Peter R. Saulson, {\em Fundamentals of Interferometric Gravitational Wave Detectors} (World Scientific, Singapore, 2017). 
\bibitem{Melissinos2010} Adrian Melissinos and Ashok Das, ``The response of laser interferometers to a gravitational wave,'' Am. J. Phys. {\bf 78}, {1160--1164} (2010). \doi{10.1119/1.3443566} 
\bibitem{Ruggiero2021} Matteo L. Ruggiero, ``Gravitational waves physics using Fermi coordinates: A new teaching perspective,'' Am. J. Phys. {\bf 89}, 639--646. \doi{10.1119/10.0003513}
\bibitem{Cooperstock1993}  F. I. {Cooperstock}, and Valerio {Faraoni}, ``Laser-interferometric detectors of gravitational waves.,' Class. Quantum Grav. {\bf 10}, 1189--1199 (1993). \doi{10.1088/0264-9381/10/6/016} 
\bibitem{Finn2009} Lee Samuel Finn, ``Response of interferometric gravitational wave detectors,'' Phys. Rev. {\bf D79}, 022002 (2009). \doi{10.1103/PhysRevD.79.022002}
\bibitem{Koop2014} Michael J. Koop and Lee Samuel Finn, ``Physical response of light-time gravitational wave detectors,'' Phys. Rev. {\bf D90}, 062002 (2014). \doi{10.1103/PhysRevD.90.062002}
\bibitem{Congedo2017} Giuseppe Congedo, ``Detection principle of gravitational wave detectors,'' Int. J. Mod. Phys. D {\bf 26},  {1741022} (2017). \doi{10.1142/S021827181741022X}
\bibitem{Lobato2021} Jo\~{a}o C. Lobato et al., ``Influence of gravitational waves upon light in the Minkowski background: From null geodesics to interferometry,'' Phys. Rev. {\bf D104}, 024024 (2021). \doi{10.1103/PhysRevD.104.024024}
\bibitem{LISA-TDI} Massimo Tinto, Sanjeev V. Dhurandar, ``Time-Delay Interferometry,'' Liv. Rev. Relativity {\bf 8}, 4 (2005). \doi{10.12942/lrr-2005-4}
\end{thebibliography}
\end{document}